# A new launch pad failure mode: Analysis of fine particles from the launch of the first Starship orbital test flight

B. Dotson[1], P. Metzger[1], J. Hafner[2], A. Shackelford[1], K. Birkenfeld[1], D. Britt[1], A. Ford[3], R. Truscott[4], S. Truscott[4], J. Zavaleta[4], J. Zemke[5], K. Purvis[6], M. Scudder[7], C. Johnson[8], J. Galloway[9], J. DeShetler[9]

[1]University of Central Florida, Department of Physics, 4000 Central Florida Boulevard, Orlando, FL 32816
[2]Rice University, Department of Physics and Astronomy, 6100 Main St, Houston, TX 77005
[3]Independent Researcher, Houston, TX 77005
[4]Independent Researcher, Port Isabel, TX 78578
[5]Independent Researcher, Portland, OR 97212
[6]Independent Researcher, Bella Vista, AR 72715
[7]Independent Researcher, Centerville, MA 02632
[8]Independent Researcher, Homewood, AL 35209
[9]Independent Researcher, Charlotte, NC 28278

**1.0 ABSTRACT:**
This study examines the characteristics, composition, and origin of fine particle debris samples collected following the launch of the first Starship orbital test flight, which suggests a new launch pad failure mode previously unknown. Particle shapes, sizes, bulk densities, and VIS/NIR/MIR spectra, of collected fine particle material from Port Isabel, TX, were analyzed and compared to pulverized concrete, Fondag (high temperature concrete), limestone, and sand recovered from the area near the Starship launch pad after this test flight. Raman spectroscopy was also used to determine mineral compositions of each sample. Results suggest that the fine particle material lofted by the Starship launch is consistent with sand derived from the launch site. These results imply that the destruction of the launch pad eroded and lofted material into the air from the underlying sandy, base-layer. From calculations, this lofted material likely remained suspended in the air for minutes following the launch from recirculation, allowing for transport over an extended range. Most of the recovered material was too coarse to be a respiration hazard, as a small mass fraction of the particles (<1%) had diameters of 10 um or less. Video analysis and ballistic models also provide insight into the failure mechanism associated with the launch pad, which was consistent with a high-pressure eruption from the region below the failed launch pad. As one of the vehicles selected for NASA's Human Landing System (HLS) contract, the results of this study clearly highlight the implications of plume effects and pad designs for future launches from Starbase, TX, as well as for NASA's Artemis program.

**2.0 INTRODUCTION:**
On April 20, 2023, SpaceX conducted the first orbital test flight attempt of Starship, a prototype super-heavy rocket, from the SpaceX Starbase launch site in Boca Chica, TX. The first stage of this rocket, with 33 liquid oxygen and liquid methane Raptor engines, produced roughly 18 million pounds of thrust at lift-off (Seedhouse, 2022). Without a water deluge system or flame diverter to help mitigate the effects of rocket exhaust, the Starship pad suffered major damage during the powerful launch, creating a large crater underneath the Starship and throwing large pieces of concrete and debris hundreds of meters (Hull & Grush, 2023). Within minutes of the Starship launch, residents of Port Isabel, TX, roughly 10 km (5 -6 miles) from the Boca Chica launch site, also reported that wet, sandy, fine particle material of unknown origin fell from the sky onto populated areas. Some of these dislodged concrete, Fondag (high temperature concrete), and fine particle samples were collected by private citizens and sent to the University of Central Florida and Rice University for analysis to determine the composition and potential origin of the granular material that fell on Port Isabel, TX, directly following the orbital test flight.

## 3.0 METHODS:

A total of four fine particle samples, having reportedly rained down on the surrounding areas following the first Starship launch, were collected at different locations near Port Isabel, TX by private citizens to support this study. These samples were taken from surfaces such as the hood of a car or from patio furniture within several hours of the Starship test flight and at a straight-line distance between roughly 8.47-10.73 km (5.26 – 6.67 miles) from the launch site as shown in Fig. 1. Of note: one of the collected fine particle samples of interest, Sample A, was taken from the hood of a car parked 10.73 km (6.67 miles) away roughly 6 days after the owner drove several hours to Houston, TX. Larger pieces of the Fondag and concrete material from the launch pad, as well as a large piece of limestone, were also recovered by private citizens on the beaches immediately adjacent to the Starbase launch site as well. The large piece of limestone recovered from Boca Chica Beach, TX, also contained some amount of beach sand lodged in cracks of the limestone sample. This sand was recovered by rinsing it from the visible cracks with Isopropyl Alcohol and then baked at 140°C for 20 mins.

The collected fine particle samples were transferred into cylindrical containers with a known volume for analysis. The mass of each sample was measured using a BSM220.4 electronic balance, with a published error of ±0.1 mg. The volume of each sample was estimated using digital calipers, with a published error or ±0.01 mm, to measure the material's height within the cylindrical container. Average bulk density of each sample was then calculated using the sample mass and volume measurements.

Particle shapes and size distributions, including average particle size, particle size distribution parameters (D10, D50, and D90), average aspect ratio, elongation, and sphericity, were measured for each collected sample using a Cilas-1190 connected to a CETI Inverso TC 100, inverted optical microscope with an objective magnification of 4x and ExpertShape analysis software. This system uses an IDS uEye UT149xLE-C camera to digitally record microscope images with a numerical aperture of 0.10 and a resolution of 2.452um. For all images analyzed, a threshold value of 58% was used with a total particle number (N) greater than 648 particles of various sizes (with the exception of uncrushed Fondag).

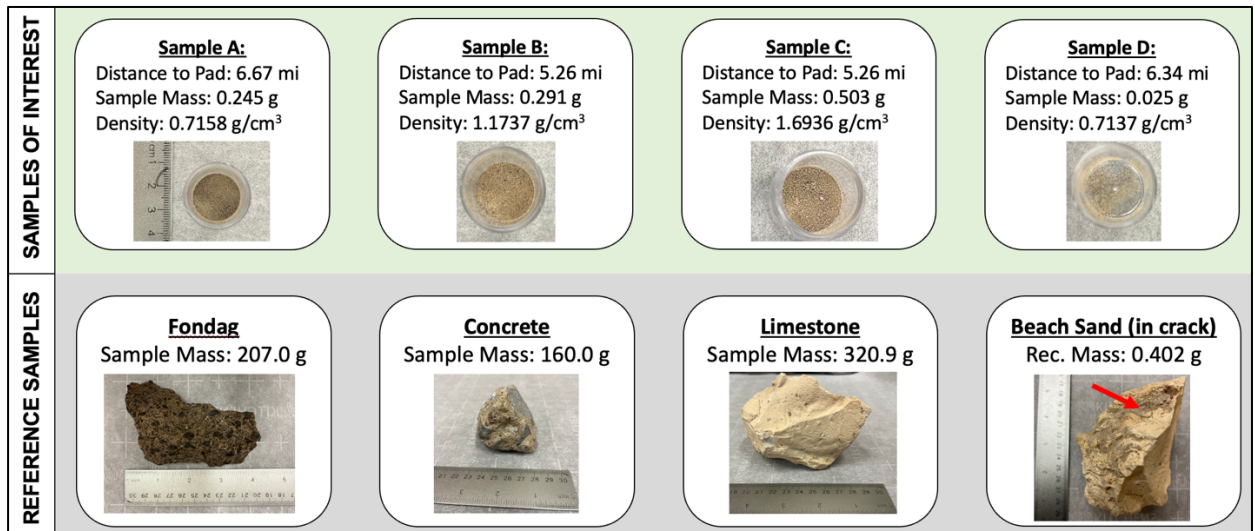

Figure 1. Collected samples following Starship orbital test flight.

The VIS/NIR and MIR spectra from each recovered sample was measured using a Nicolet iS50 FT-IR spectrometer with a DTGS KBr Detector and beam splitter, as well as Si beam splitter and InGsAsfor visible spectra. Prior to measuring spectra for each sample, background spectra were recorded to account for atmospheric water effects. Measurements were taken using the Pike EasiDiff reflectance accessory for the examined samples and the ambient background. For the larger recovered samples of Fondag, concrete, and limestone, small portions were removed using a rock hammer, and crushed manually using a clean mortar and pedestal until the particle sizes were comparable to those of the collected fine particle samples, since particle size can impact the width and depth of features in MIR reflectance spectra (Udvardi et al., 2017). A total of three spectra were acquired for each sample, rotating the sample cup for each measurement to account for any specific particle sizes or surface features. These spectra were then averaged and normalized to features at 4000 cm$^{-1}$ for all samples.

Raman spectra were recorded with an Ocean Insight fiber optic Raman probe and 785 nm stabilized diode laser. The laser power was adjusted so that 75 mW is emitted from the probe, which focuses the light with a 0.22 NA lens. The spectrometer was calibrated for wavelengths with a xenon calibration lamp, and for spectral efficiency with an incandescent quartz tungsten halogen lamp. The probe was mounted to a motorized X-Y-Z translation stage and scanned across the sample surface, either that of a solid rock or of dust in the bottom of a plastic dish. Raman spectra were recorded at 0.05 mm lateral resolution and the probe height was adjusted in Z at each point to maximize the Raman signal. For each sample, 400 spectra were recorded over a ~1 mm$^2$ area.

Video and images of the Starship launch were collected by both SpaceX and private citizens from multiple angles and distances relative to the launch pad. These videos and images, when coupled with first-hand accounts of the resulting precipitation and debris, were then used to guide initial ballistics calculations for ejected velocities in Mathematica, as well as debris cloud heights and shapes. Objects of known size or distance, including the Starship vehicle itself, were used to calibrate the videos and images. Measured particle shapes, bulk densities, and aspect ratios from collected samples were combined with ballistic models to investigate mechanisms associated with particle transport and launch pad failure as described in Sections 5.4 and 5.5 below.

**4.0 RESULTS:**
**4.1 Physical Measurements:**

A comparison of the collected samples of interest, fine particle material sampled at various distances from launch pad, is shown in Fig. 1 and Table 1, compared to larger reference samples of Fondag, concrete, and limestone with recovered beach sand. As shown, the amount of recovered fine particle ranged between 25.2 – 245.1 mg, while bulk densities ranged from approximately 0.71 – 1.69 g/cm$^3$ for the samples of interest. Of note, Sample D contained very little sample when compared to other collected samples.

The average particle size, as well as size distribution parameters (D10, D50, and D90) and particle shapes are reported for each examined sample in Table 1. As discussed above, the reported size and shape values for Fondag and concrete in Table 1 are from small pieces that naturally fractured from the full reference sample with a rock hammer. The values reported for crushed Fondag and concrete represent the resultant material after using a mortar and pedestal as described above. Of note, the amount of fractured Fondag, recovered after hitting the larger reference sample with a hammer only contained 98 observed particles, generally larger in size (average particle size of 693.66 um).

Table 1. Size and shape data for examined samples.

| Specimen | Number of Particles Examined | Average Particle Size (µm) | D10 | D50 | D90 | Sphericity | Aspect Ratio | Elongation |
|---|---|---|---|---|---|---|---|---|
| Sample A | 694 | 83.57 | 46.2 | 86.26 | 169.73 | 0.53327 | 0.65462 | 0.34538 |
| Sample B | 708 | 193.37 | 124.88 | 190.3 | 325.45 | 0.60227 | 0.6777 | 0.3223 |
| Sample C | 692 | 233.03 | 142.64 | 231.65 | 370.82 | 0.59909 | 0.68274 | 0.31726 |
| Sample D | 722 | 323.59 | 199.73 | 310.46 | 609.67 | 0.55652 | 0.66519 | 0.33481 |
| Fondag | 98* | 693.66 | 410.83 | 767.22 | 1049.77 | 0.55125 | 0.63231 | 0.36769 |
| Crushed Fondag | 801 | 255.99 | 114.6 | 289 | 421.71 | 0.62225 | 0.682 | 0.318 |
| Concrete | 649 | 171.57 | 91.72 | 188.66 | 279.92 | 0.61656 | 0.67682 | 0.32318 |
| Crushed Concrete | 721 | 320.38 | 163.28 | 302.18 | 582.47 | 0.56163 | 0.66281 | 0.33719 |
| Limestone | 714 | 167.69 | 100.91 | 195.34 | 231.36 | 0.7454 | 0.7293 | 0.33719 |
| Beach Sand | 661 | 236.18 | 145.83 | 242.23 | 383.35 | 0.62576 | 0.68824 | 0.33719 |

*Uncrushed sample contained a limited amount of particles*

Shown in Table 1, the average particle sizes for Samples A-D ranged from 83.57 - 323.59 um, with less than 10% of a sample having particle sizes greater than 609.67um for any sample (D90 coefficient). Direct measurements of particle size distributions revealed that while Samples A-D contained some number of particles less than 10um in diameter, the amount was less than 1% for all particles examined in Samples A-D. Measurements of particle shapes revealed average aspect ratios that ranged from 0.65 – 0.68, and elongation parameters measured between 0.32 – 0.35. Combined with sphericity measurements, this implies that recovered particles were not perfectly spherical and generally had some amount of elongation; These shapes will have an impact on the drag profile for particle calculations.

**4.2 VIS/NIR and MIR Spectroscopy:**

In comparing measured VIS/NIR spectra between wavenumbers of 9,000 - 15,000 $cm^{-1}$ for all samples, there were no meaningful spectral features observed that would allow for comparison in this spectral region. These VIR/NIR spectra demonstrated a broad maximum reflectance peak near 12,200 $cm^{-1}$, that gradually decreased by roughly 60% at 9,000 $cm^{-1}$ and roughly 20% at 15,000 $cm^{-1}$, for all samples examined. Aside from being generally featureless, these VIS/NIR spectra otherwise only demonstrate a potential for similar compositions between materials but requires other wavelength regions for further diagnostics.

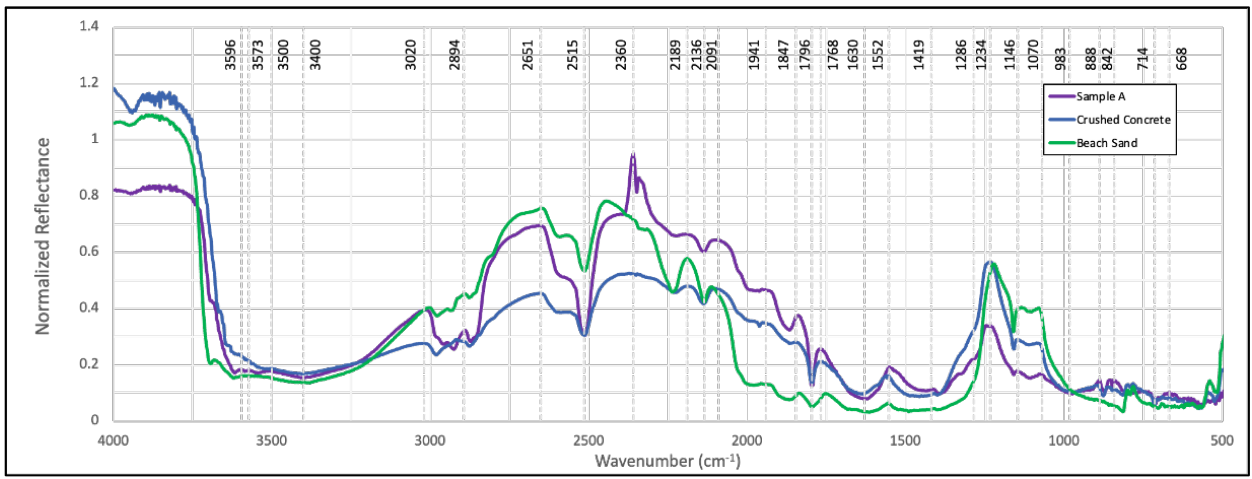

Figure 2. MIR spectra for Sample A compared to beach sand and concrete

The MIR spectra for Sample A compared to recovered beach sand and crushed concrete is presented in Fig. 2 above. Similarly, the MIR spectra for Samples B-D are shown compared to recovered beach sand in Fig. 3 as well. In both Fig. 2 and 3, vertical gray lines mark wavenumbers for spectral features of interest. From the spectra, there is a board reflectance spectral feature between roughly 2,000 – 3,400 $cm^{-1}$, and a prominent doublet spectral feature between roughly 983 - 1,286 $cm^{-1}$ as well. These same spectral features are observed in the MIR spectra for concrete (Fig. 2), except the feature extends over a larger range of wavenumbers from 1,630 – 3,400 $cm^{-1}$ for concrete. Also of note, a singlet or doublet spectral feature was observed at 2,360 $cm^{-1}$ that changed between spectra (even for the same sample) during the measurement. These specific features at 2,360 $cm^{-1}$ appeared to be largest after opening the FTIR device for each sample and decreased in normal reflectance gradually over roughly 3-5 minutes.

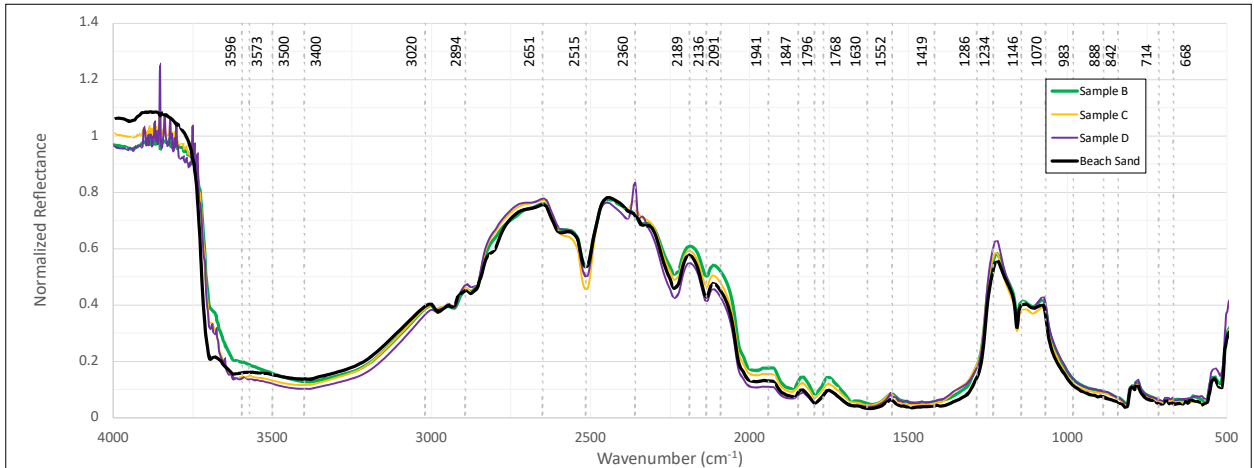

Figure 3. MIR spectra for beach sand compared to Samples B, C, and D.

### 4.3 Raman Spectroscopy:

Raman spectral maps were recorded across Sample A-D as well as the concrete, Fondag, limestone, and beach sand. Every spectrum in a map was analyzed and compared to quartz, calcite, and hematite spectra in RRUFF™ (not an acronym; proprietary Raman database (Lafuente et al., 2015)). The individual mineral peaks from the database were then fit to the corresponding regions in the experimental spectra to get an amplitude for each peak, as seen in Fig. 4. At each position, if the measured amplitude ratios were similar to a mineral's ratios from the database spectra, the location was counted as including that mineral, and the amplitude of the strongest peak represents the mineral intensity.

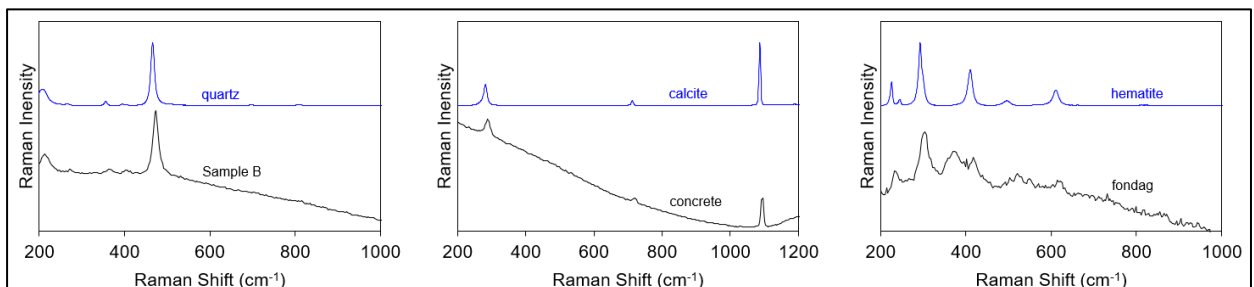

Figure 4. Raman spectra from Sample B (left), concrete (middle), and fondag (right), including matching reference spectra from the RRUFF™ database.

The mineral intensities were summed for all pixels in a sample to represent the total content of each mineral that is visible in Raman spectra. The total contents were used to calculate the quartz/calcite ratios in Table 2, which is a good indicator for sand (predominantly quartz) or concrete (predominantly calcite). Spectra that resemble calcium aluminates were also detected, but no RRUFF™ entries were available.

Table 2. Quartz-to-Calcite ratios from Raman spectroscopy.

| Specimen | Quartz : Calcite | Other Minerals Present |
|---|---|---|
| Sample A | -- | Organic compounds (Likely contamination) |
| Sample B | 24.555 | |
| Sample C | 26.728 | |
| Sample D | 12.805* | |
| Fondag | 0.038 | Hematite, Calcium Aluminates |
| Crushed Fondag | 0.000 | Hematite, Calcium Aluminates |
| Concrete | 0.150 | |
| Crushed Concrete | 0.355 | |
| Limestone | 0.026 | |
| Beach Sand | 8.843 | |

*Sample contained a limited amount of particles.

**4.4 Visual Observations of the Event:**

The Starship engines ignited about 2 seconds before liftoff. Videography looking down from above the launch pad shows the plume was divided into six radially outward lobes, divided by the six legs of the Orbital Launch Mount, giving the appearance of a six-petaled flower. All six lobes curve upward and reached altitudes on the order of 100 m at their outer edges (away from pad centerline) by about T+5 seconds, with two of the northerly lobes curving upwards higher due to infrastructure blockages at ground level. At T+5 seconds several large pieces of debris suddenly appeared in the center of this "flower", traveling upward near the rocket. A seventh plume lobe also suddenly appears at T+5 to 6 seconds near the center of the flower but offset slightly toward the south. This seventh lobe expanded vertically instead of radially like the other six, and it grew much faster than the others, rapidly catching them in height. Our hypothesis is that this seventh lobe was the result of the launch pad rupturing.

Larger debris the sizes of rocks and boulders fell in a local debris field as noted after the launch. Rocks traveling radially outward close to the ground were at high velocity. Splashes as high as 20 m (the height of a six-story house) were observed in the ocean near the launch pad, presumably caused by very large debris that had traveled in a higher ballistic trajectory. Imagery later than the initial pad failure is focused mostly on the Starship high in the sky, so there is a short gap in the publicly available imagery of the pad.

At T+23 seconds, long-range imagery from the west shows that the overall plume (the six petals plus the central seventh) had reached a height of about 260 m, using the Starship as a reference scale. Videography looking from Port Isabel can be used to calculate the distance and speed of the approaching cloud. By then it had a large gap between its bottom and the ground. This indicates it had continued to rise during the 17 second gap of imagery, possibly due to thermal buoyancy of the hot plume gas, or that the bottom had been sheared off by the winds (discussed below). It is possible that thermal convection also drove rotational motion inside the cloud that kept the debris suspended longer (discussed below).

Still at a distance, before the cloud reached Port Isabel, precipitation was imaged falling from the cloud (Fig. 5). At T+13 minutes 1 second as the cloud moved overhead, some coauthors (R. and S. Truscott) observed that it started raining at their positions; noting that it was sand falling from the sky although it felt wet and was mixed with rain (Fig. 5). Another coauthor (Ford) located near the first observers, noted that the sand felt wet and made dark circles on clothing reminiscent of raindrops absorbing into the cloth, but it was completely dry when subsequently touched and could be brushed off as completely dry sand. Imagery of the sand stuck to a laptop computer screen shows it was arranged in diagonal streaks running down the screen, such that a series of smaller grains led diagonally down to a larger grain at the termination of each streak, so it was reminiscent of raindrops running down glass leaving a trail of smaller droplets. This suggests the sand grains were enclosed in raindrops. Another coauthor (Zavaleta) observed that at a location slightly farther north, at a distance of 10.32 km (6.34 miles), the sand reached ground level without these indications of moisture. We hypothesize the cloud was indeed raining both water and particulates, but the water quickly evaporated since the local weather conditions were not favorable for rain, and the rain did not reach ground level at all for the locations farther from the pad.

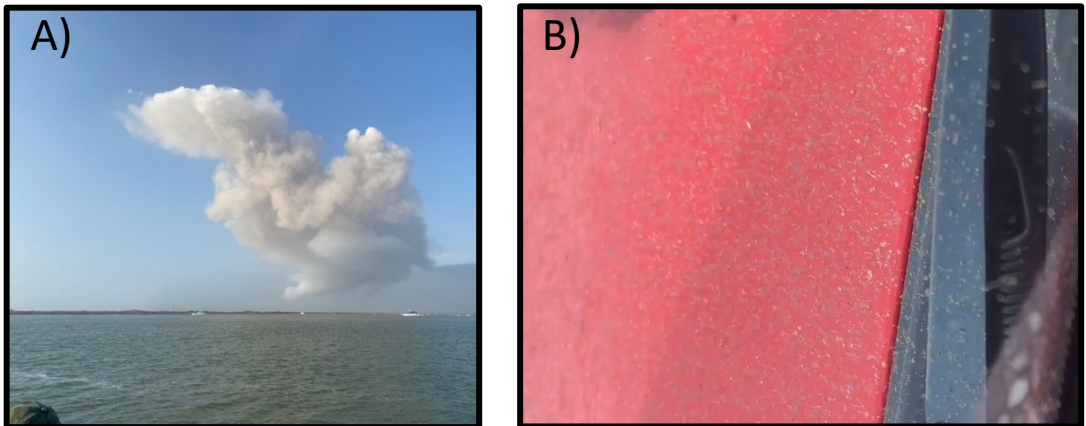

Figure 5. Images of A) post-launch cloud traveling towards Port Isabel and B) Falling particle materials on the hood of a car 8.47km (5.26 miles) away from launch pad.

## 5.0 DISCUSSION:
### 5.1 Physical Measurements:

Based on initial measurements, the amount of collected fine particles at various locations, as well as the estimated bulk densities, varied based on distance from the launch pad. Generally, fine particle samples with bulk densities less than 1.0 g/cm$^3$ (Samples A and D) were collected farther from the launch pad, while more dense samples (Samples B and C) traveled shorter distances. Bulk density measurements may not be consistent with actual material densities at the collection site, since the mechanism by which the samples were collected, transported, and shipped to the lab could ultimately effect the bulk density. Nevertheless, the average bulk density for all samples of interest provides a rough estimate for ballistic calculations (discussed in Section 5.4 and 5.5).

All fine particle samples had roughly the same shape and degree of elongation as the reference materials, including particles of beach sand. Such particle elongation would yield a higher drag coefficient for particles, when compared to spherical particles, as they were lofted and transported through air. A higher drag coefficient would therefore cause shorter trajectories when considering ballistic calculations in atmosphere, as discussed below.

As presented in Section 4.1, the average particle sizes and shapes for Samples B-D generally matched those for beach sand, while Sample A had a slightly smaller average particle size. Given that Sample A was collected the furthest from the launch pad from all other samples, these measurements potentially imply a ballistic transport mechanism where smaller particles generally covered longer distances from the pad. Nevertheless, since less than 1% of all examined samples contained particles less than 10 um, it is believed there is little negative health or environmental impacts from the fine particle material. Natural storms or wind processes can routinely transport material with similar particle sizes as observed in this study.

**5.2 VIS/NIR and MIR Spectroscopy:**

While the VIS/NIR spectra of all particles did not provide sufficient spectral features to fully characterize samples, Samples B-D have MIR reflectance spectra consistent with beach sand. Samples B-D are different from the other reference samples, including crushed Fondag, crushed concrete, and crushed limestone. The MIR spectra for Sample A are slightly different from the other fine particle samples collected. The overall composition of Sample A cannot be determined from VIS/NIR/MIR spectra alone, but the material generally has MIR spectral features consistent with either beach sand or potentially crushed concrete. However, it is important to note that smaller grained sands often acts as a filler material in concrete, causing their MIR spectral features to be roughly similar. Additionally, Sample A also likely had some contamination as a result of being collected from the hood of a car, roughly 6 days after the launch.

The singlet and doublet spectral features near 2,360 $cm^{-1}$ are likely associated with atmospheric water, as these features changed while measuring spectra for a given sample, ultimately decreasing over time as spectrometer was slowly purged of air. Thus, differences in these spectral features between samples near 2,360 $cm^{-1}$ can likely be ignored.

**5.3 Raman Spectroscopy:**

Raman spectroscopy can detect low frequency vibrations with sharp spectral peaks so it provides good specificity in the detection of different minerals. Here the Raman quartz/calcite ratio from peaks at 460 $cm^{-1}$ and 1200 $cm^{-1}$, respectively, is used to associate the samples with sand (quartz) or concrete (calcite). Table 2 shows that samples B-D are made up predominantly of quartz and therefore represent sand blown from the launch site rather than concrete. Sharp Raman peaks were not detected in sample A, consistent with its potential contamination.

**5.4 Video Analysis and Ballistics Calculations:**

Based on video analysis and observed debris, it is believed that the seventh lobe noted during the initial launch sequence was the result of the launch pad rupturing. The offset from centerline matches our expectations since the Starship's engines were continuing to thrust downward on centerline preventing an upward eruption from climbing higher on center and because the pressure imbalance on the pad (discussed below) would have been greatest off centerline. The timing of the seventh lobe and debris therefore fixes the time of pad failure to about T+5 seconds.

The anomalous rain can be explained because the cloud was composed largely of water vapor from the rocket exhaust, and drops were seeded in this cloud by the presence of particulates from the pad failure. The particles might have stayed aloft even longer than they did due to the cloud's internal convection, and therefore would not have fallen densely on Port Isabel, but we hypothesize that the water accreting onto the particulates increased their mass and ballistic coefficient, so they precipitated. The fall of sand mixed in rain lasted less than 1 minute in the locations where the authors were monitoring. The sand in several locations achieved typical surface densities as shown in Fig. 1, but this was not accurately measured.

The authors noted that this falling sand began only when the plume cloud had moved directly overhead, so the sand was correlated to the cloud rather than traveling ballistically from the launch pad outside of any cloud. The distance from the launch pad to the first location of observed falling sand in Port Isabel was 9.82 km (6.10 miles), indicating the cloud had traveled with an average speed of 12.5 m/s (28 mph) to the northwest. Wind data show that this cloud speed is consistent with the wind speed and direction for altitudes 250 m and higher. Lower altitudes had slower and more northerly winds, yet the cloud approaching Port Isabel was not visibly shearing except perhaps in the distance near the launch pad, so this constrains the altitude of the cloud's bottom to at least 250 m near Port Isabel. It is possible that shearing below 250 m dispersed the lower part of the cloud early, or it is possible the entire cloud rose thermally to above 250 m. The authors observed that the precipitation lasted less than one minute. With the cloud moving over an observation site at 12.5 m/s (28 mph), the precipitation would last less than 1 minute as observed if the cloud's lateral dimension was about 750 m. Using this as a length scale for the videography and assuming the in-plane and out-of-plane lateral dimensions are equal, the cloud bottom is roughly estimated (order of magnitude) to have been 250 to 300 m altitude and the cloud's top 450 to 500 m.

Initially we sought to explain the particle trajectories as purely ballistic. An algorithm for Newtonian integration of their trajectories using the drag equation (Equation 1 below) was performed, where the Reynolds number $\text{Re}_{\text{rel}} = \rho D |\vec{v} - \vec{u}|/\nu$, $\rho$ = air density, $D$ = particulate diameter, $\vec{v}$ = particulate velocity, $\vec{u}$ = local wind velocity, $\nu$ = air viscosity, Knudsen number $\text{Kn} = \lambda/D$, and $\lambda$ = mean free path length of the gas molecules.

$$F_{\text{drag}} = \frac{24}{\text{Re}_{\text{rel}}} \cdot \frac{1 + 0.15\,\text{Re}_{\text{rel}}^{0.687}}{1 + \text{Kn}\cdot[3.82 + 1.28\,\text{Exp}(-1/\text{Kn})]} \qquad (1)$$

This analysis found that sand-sized particles stop at distances that are very short relative to the 9.8 km (6.1 mile) downrange transport, so to good approximation the horizontal and vertical equations of motion can be decoupled. Winds and observations indicate 13 minutes are needed to reach Port Isabel in the horizontal direction. Balancing drag with gravitational force, the terminal velocities in the vertical direction as a function of particle size were found and the starting heights to produce 13 minute freefall are shown in Fig. 6. The necessary height for 300 to 500 $\mu$m particles would be 2-3 km, which is excessive, so we reject the hypothesis that transport was ballistic. This indicates there was circulation in the cloud (Blyth et al., 2005) that kept particles suspended as it moved downrange. Further analysis of the cloud dynamics are not necessary to analyze the launch pad failure. The salient point is that the initial burgeoning of the seventh plume lobe reached 260 m in a few seconds, climbing higher as the cloud reached our observation sites, and that subsequent transport was due to atmospheric dynamics.

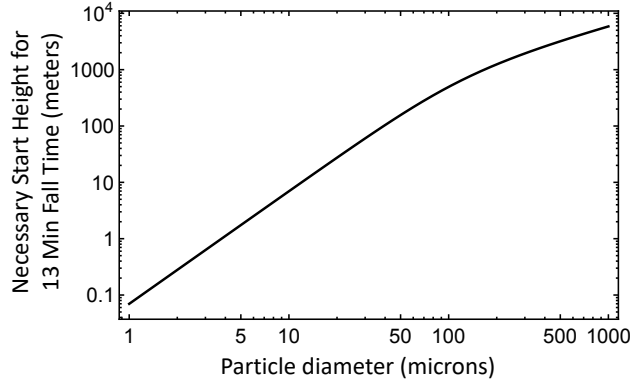

Figure 6. Initial heights of particles needed for a 13 minute fall time, in still air (neglecting wind circulation in the cloud).

**5.5 Pad Failure Model**

We hypothesize that the launch pad failed as follows: application of mechanical stress and thermal stress resulted in concrete fractures that grew until they spanned the pad's thickness, allowing hot, high pressure gas to penetrate beneath its surface. The shallow groundwater was vaporized by the hot gas, lowering the temperature but increasing the pressure. If the pressure under the pad was initially equilibrated to the shock stagnation pressure, which we approximate as the dynamic pressure of the Starship Raptor engines (thrust divided by nozzle exit area), this was 2.6 MPa, increasing with groundwater vaporization. As the Starship lifted off, the plumes would mix with turbulent ambient gas reducing impingement pressure on the pad, which would further exaggerate the pressure imbalance above versus below the concrete until failure occurred.

Chunks of ejected concrete were found as far away as 817 m (2,680 feet) (Hull, Grush, and Leopold, 2023). Since the most distant debris would have been those with a 45 deg ejection angle, the initial velocity is calculated at 90 m/s. Impact splashes in the ocean up to 20 m high, several hundred meters from the launch pad, were seen in videography 9 – 12 s after liftoff. Assuming 45 deg ejection angle for the chunks that caused those splashes predicts initial velocity of 62 – 83 m/s. Assuming 90 m/s to match the most distant chunks predicts ejection angle 29 – 41 deg. Lower angle but faster debris would have reached the ocean first, so the splashes corroborate the estimate of 90 m/s initial ejection velocity.

Publicly available imagery of the crater under the pad after the launch suggests on order of 1/3 of the crater's volume was into the sand below the concrete (Fig. 7) whereas the other ~2/3 was composed of the overlying layers of concrete construction, both ordinary and refractory concrete (Fondag). Some of the sand would have been removed in the initial blast and some due to continuous erosion, from below the launch pad (Fig. 7), by the Starship's ongoing plume. To bound the problem, we assume either 0% or 100% of the final volume of sand was ejected with the initial blast. Using 100%, assuming the sand had porosity $n = 0.3$ saturated with water that vaporized completely, the mass of gas contributing to the explosion was 4.2%wt of the mass of ejected concrete. Assuming 0% of the sand was removed in the initial blast, this would be ~0%wt. We compare this to volcanic explosions where a caprock over pressurized volatiles ruptures and fragments. This impulsive explosion model is an idealization because as the pad fractured, the Starship would be continuously sourcing hot gas that would continue vaporizing groundwater and excavating the crater, mixing more material into the initially impulsive cloud, but the initial blast would produce higher velocity ejecta to compare with the debris velocity calculations.

Fagents and Wilson (1993) simulated the physics of volcanic eruption and found that a 2.6 to 10 MPa source of gas (matching Starship shock recovery pressure with some mass of vaporized groundwater) with mass fraction of the gas between 1%wt and 10%wt of the ejected rock would produce an initial ejection velocity of 90 m/s to 370 m/s for the debris. The velocity is not very sensitive on the pressure across this range but varies highly with mass fraction of the gas. The lower end of the range matches the observed ejecta velocity, implying mass fraction of the gas was ~ 1%wt. Solving for the volume of eroded sand with groundwater indicates 8% of the volume of the initial blast crater was into the sand, the other 92%wt being the concrete. This indicates pad failure occurred with only limited, but non-zero groundwater vaporization, which seems reasonable since heat transfer through the wet sand would be limited by thermal conductivity such that only the upper portion of the wet sand could participate in the initial blast. The volume of removed sand would then have increased a factor of 3.25 after the initial blast to achieve the final observed crater size, contributing more sand and water vapor as a continuous process behind the initial ejecta wave.

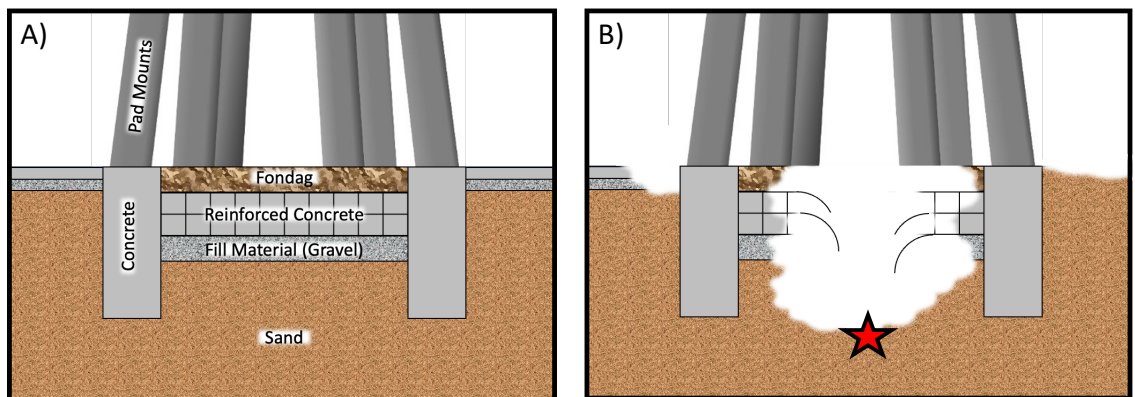

Figure 7. Pad before and after failure (A and B, respectively). Potential source of sand collected in Port Isabel, TX, is highlighted by the red star [Not to scale].

**5.6 Application to Lunar Launch and Landing Pads**

This model matches all observations using only reasonable assumptions. We conclude that this is a reasonable model of the failure: concrete erupting under tension because of buildup of gas beneath its surface, comparable to volcanic caprock failure, followed by continuous sand erosion by the rocket plume after the initial explosion of the pad. The primary driver for this failure is the tension produced by the injected plume gases, somewhat enhanced by vaporized groundwater. Identifying this failure mode is important as we go forward to design lunar landing pads. For lunar construction, the cost of transporting construction equipment and the use of energy in construction, not to mention the time-value of completing the construction, are extremely high. So the lunar exploration community has considered only minimalist construction techniques to reduce these costs. These techniques include sintering a flat surface with either microwaves or solar concentrators, application of polymers brought from Earth as a cement, or development of waterless "lunacrete" that cures in vacuum (Metzger & Autry, 2023). Some of these methods may be at greater risk to thermal fracture, pressure buildup beneath the surface, and explosion. Ejected concrete debris on the Moon would be a risk to a launching or landing vehicle as well as surrounding assets on the lunar surface, although the ejection velocities might be less than in the Starship launch event due to the absence of groundwater. Further analysis of the failure mode with high fidelity computational fluid dynamics (CFD) modeling is urgently needed to quantify the lunar case. Nevertheless, the results of this study help to identify and constrain a previously unrecognized failure mode that could occur on the Moon, which might result in loss of vehicle and crew (or loss of an uncrewed mission) unless modeled and mitigated adequately.

## 6.0 CONCLUSIONS:

Fine particle samples collected roughly 10 km (6.2 miles) away from the Starship launch pad have measured particle shapes, sizes, spectra, and mineral compositions that are consistent with beach sand, not pulverized concrete or Fondag from the pad itself. This beach sand was likely lofted from beneath the launch pad during launch, as supported by ballistic calculations and models. With few particles less than 10um and lofting duration of 1-2 minutes, beach sand lofted and transported by this launch likely had little environmental or health impacts to the neighboring area, as similar material can frequently be lofted by natural events like storms or wind. Nevertheless, because of damage at the launch site, improvements to the launch pad design for this class of rocket are strongly recommended prior to future launches. While installing dust sensors near the launch pad may also provide additional information on material transport processes, improvements to the launch pad design to prevent explosive pad failures are preferable to reduce the frequency of such transport phenomenon. Video analysis of the launch sequence, coupled with ballistic models, suggest a potential failure mechanism of the launch pad consistent with a high pressure volcanic-like eruption, with penetration of gas through fractures to the underside of the pad as the most likely cause of failure. The presence of vaporized groundwater only enhanced the already major gas penetration and eruption phenomenon. Understanding the failure mechanism of launch pads, as well as the ballistic and transport processes for debris, will be crucial for future pad designs. Future studies should use the results of this study to estimate particle transport phenomena and cratering for a similar super-heavy class rocket on the Moon, in preparation for NASA's Artemis landings. While the gravity and atmospheric conditions are much different on the Moon, making it easier for lofted material to travel long distances, the results of this study point to a new pad failure mode that could occur while launching and landing a rocket of this size on the Moon, even with a prepared pad.

## 7.0 REFERENCES:

Blyth, A. M., Lasher-Trapp, S. G., & Cooper, W. A. (2005). A study of thermals in cumulus clouds. *Quarterly Journal of the Royal Meteorological Society*, *131*(607). https://doi.org/10.1256/qj.03.180

Fagents, S. A., & Wilson, L. (1993). Explosive volcanic eruptions—VII. The ranges of pyroclasts ejected in transient volcanic explosions. *Geophysical Journal International*, *113*(2). https://doi.org/10.1111/j.1365-246X.1993.tb00892.x

Hull, D., & Grush, L. (2023, May 1). SpaceX Starship rocket launch hastily approved, suit against FAA says. *The Seattle Times*, 1–5.

Lafuente, B., Downs, R. T., Yang, H., & Stone, N. (2015). RRUFF$^{TM}$ Project. In *The power of databases: the RRUFF project. Highlights in Mineralogical Crystallography, T Armbruster and R M Danisi, eds.*

Metzger, P. T., & Autry, G. W. (2023). The Cost of Lunar Landing Pads with a Trade Study of Construction Methods. *New Space*, *11*(2). https://doi.org/10.1089/space.2022.0015

Seedhouse, E. (2022). Starship. In *SpaceX*. https://doi.org/10.1007/978-3-030-99181-4_9

Udvardi, B., Kovács, I. J., Fancsik, T., Kónya, P., Bátori, M., Stercel, F., Falus, G., & Szalai, Z. (2017). Effects of Particle Size on the Attenuated Total Reflection Spectrum of Minerals. *Applied Spectroscopy*, *71*(6). https://doi.org/10.1177/0003702816670914